\documentclass{aa} 
\usepackage{psfig} 
\usepackage{graphics} 
\usepackage[T1]{fontenc}
\usepackage{times}

\newcommand{\abbrev}[1]{{ #1}} 
 
\newcommand{\cdf}{{\sc cdf}}

\newcommand{\oldhb}{HB$_{\rm OLD}$}

\newcommand{\msol}{$M_\odot$}
\newcommand{\cmd}{{\sc cmd}}
                 
\newcommand{\lesssim}{\la}                     
\newcommand{\gtrsim}{\ga}                      
\newcommand{\etal}{{et al.}\ }             
\newcommand{\eg}{{e.g.},\ }                
\newcommand{\ie}{{i.e.}}                
\newcommand{\cf}{{cf.}\ }                  
\newcommand{\hi}{H{\sc i}\ }                  

\newenvironment{planotable}[1]
{\begin{table}\caption[]{\ptcaption}
\begin{flushleft}\edef\tableformat{\string#1}\ptcolsep
\begin{tabular}{\tableformat}}{\noalign{\smallskip}\hline
\noalign{\medskip}\end{tabular}
\\\ptcomments\medskip\ptrefs\end{flushleft} 
\end{table}}

\newenvironment{planotable*}[1]
{\begin{table*}\caption[]{\ptcaption}
\begin{flushleft}\edef\tableformat{\string#1}\ptcolsep
\begin{tabular}{\tableformat}}{\noalign{\smallskip}\hline
\noalign{\medskip}\end{tabular}
\\\ptcomments\medskip\ptrefs\end{flushleft}
\end{table*}}

 \def\ptcolsep{\relax}
\def\tablecaption#1{\gdef\ptcaption{#1}} \def\ptcaption{\relax}

 \def\ptcomments{\relax}

 \def\ptrefs{\relax}

\newcommand{\nl}{\\}
\newcommand{\tablehead}[1]{\hline\noalign{\smallskip}#1\\}
\newcommand{\colhead}[1]{\multicolumn{1}{c}{#1}}
\newcommand{\startdata}{\noalign{\smallskip}\hline\noalign{\smallskip}}

\def\nodata{~\dots}

\newcommand{\tablewidth}[1]{\typeout
{----- tablewidth not implemented with A\&A ----------------------}}{}
{}

\newcommand{\feh}{-1.39}
\newcommand{\corfeh}{-1.0}
\newcommand{\errfeh}{0.15}

\begin{document}

\thesaurus{03(11.06.2; 11.09.1 Fornax; 11.12.1; 11.19.5; 11.19.6)}

\title{The stellar populations of the Fornax dwarf spheroidal  galaxy. 
\thanks{Based on data collected at the European Southern 
Observatory, La Silla, Chile, Proposal N. 56.A-0538}
}
\subtitle{}

\author {Ivo Saviane \inst{1}, Enrico V. Held \inst{2}, 
and Gianpaolo Bertelli \inst{3,1} }
\offprints {E. V. Held (held@pd.astro.it)}

\institute{
Universit\`a di Padova, Dipartimento di Astronomia, 
vicolo dell'Osservatorio 5, I--35122 Padova, Italy
\and
Osservatorio Astronomico di Padova, 
vicolo dell'Osservatorio 5, I--35122 Padova, Italy
\and
Consiglio Nazionale delle Ricerche, CNR-GNA, Roma, Italy
}

\titlerunning{The stellar populations of Fornax}
\authorrunning{I. Saviane, E.V. Held \& G. Bertelli}
\date {}
\maketitle

\begin{abstract}

We present $B,V,I$\ CCD photometry of about 40000 stars in four
regions of the Fornax dwarf spheroidal galaxy down to $V \sim 23.5$,
the largest three-color data set obtained for this galaxy until now.
The resultant color-magnitude diagrams, based on a wide color
baseline, show a variety of features tracing the history of star
formation of this dwarf galaxy.
One of the most distinctive features in our diagrams is the
conspicuous young main sequence, indicating recent star formation
until approximately $2\times10^8$\ yr ago. A plume of stars brighter
than the red HB clump, with $(B- I)\sim0.5$, trace the helium-burning
phase of the young population. A comparison of the color and extension
of this feature with model isochrones suggests a relatively metal-rich
population ([Fe/H]$\sim-0.7$) with age 300--400 Myr. This represents
an important constraint for understanding the chemical enrichment
history of Fornax.
An extended upper AGB tail and a prominent red HB clump sign the
presence of the well-known dominant intermediate-age population with
an age range 2-10 Gyr, for which we have estimated a mean age
$5.4\pm1.7$.
About 0.2 mag below the red clump, an extended HB is indicative of an
old population. We show that blue HB stars may be present in the outer
regions.  Together with previous detection of RR Lyrae, this provides
evidence for a minority field population that is as old and metal-poor
as that in the Fornax globular clusters.
%
We have identified the AGB bump, a clustering of stars that occurs at
the beginning of helium shell-burning evolution, at a luminosity
$M_V\simeq-0.4$.  This is an example of the short-lived evolutionary
phases that can be revealed in stellar populations using adequately
large star data samples, whose measurements provide powerful tests of
theoretical models.

Based on precise detection of the tip of the RGB in a selected RGB
sample, we measure a corrected distance modulus $(m-M)_0 = 20.70 \pm
0.12$. An independent estimate of the distance to Fornax was also
obtained from the mean magnitude of old horizontal branch stars,
yielding a distance modulus $(m-M)_0=20.76\pm0.04$, in good agreement
with the distance estimated from the red giant branch tip and previous
results.
%
The large baseline of the $(B-I)$\ colors together with the size of
the stellar sample allowed us to analyze in detail the color
distribution of the red giant stars. We find that it can be
approximately described as the superposition of two populations. The
dominant component, comprising $\sim70$\% of the red giant stars,
consists of relatively metal-enriched intermediate-age stars. Its mean
metallicity is [Fe/H]=$\feh \pm \errfeh$, based on a comparison of the
fiducial locus of the bulk of the Fornax red giants with the
homogeneous Galactic globular cluster set of  Da Costa \& Armandroff
(\cite{daco+arma90}). Once the younger mean age of Fornax is taken
into account, our best estimate for the mean abundance of the bulk of
the galaxy is [Fe/H]$\approx \corfeh \pm \errfeh$.  The dominant
intermediate-age component has an intrinsic color dispersion $\sigma_0
(B-I) = 0.06 \pm 0.01$\ mag, corresponding to a relatively low
abundance dispersion, $\sigma_{\rm [Fe/H]} = 0.12 \pm 0.02$\ dex.
Further, there is a distinct small population of red giants on the
blue side of the RGB.  While these stars could be either old or young
red giants, we show that their spatial distribution is consistent with
the radial gradient of old horizontal branch stars, and completely
different from that of the younger population. This unambiguously
qualifies them as old and metal-poor. This result clarifies the nature
of the red giant branch of Fornax, suggesting that its exceptional
color width is due to the presence of two main populations yielding a
large abundance range ($-2.0 < {\rm [Fe/H]} < -0.7$).
This evidence suggests a scenario in which the Fornax dSph started
forming a stellar halo and its surrounding clusters together about
10--13 Gyr ago, followed by a major star formation epoch (probably
with a discontinuous rate) after several Gyr.

 
\end{abstract}

\section{Introduction} \label{sec_intro}

An increasingly large number of investigations has recognized the importance of 
dwarf spheroidal galaxies for our understanding of galaxy formation and 
evolution 
(see Mateo \cite{mate98} and  Da Costa \cite{dacosta_winter} for recent 
reviews). 
While new studies of the central, densest regions of the more distant Local 
Group 
galaxies have benefited 
from the exceptional resolution of HST, the nearby dwarf 
spheroidal satellites of the Milky Way can still be investigated in great detail 
using 
ground based wide-field data. Thanks to recent improvements in detector 
efficiency 
(in particular, in the blue part of the optical spectrum) and size, \cmd's with 
long 
color baselines and high statistical significance can be obtained in affordable 
exposure times. 

The Fornax dwarf spheroidal (\abbrev{dSph}) galaxy represents one of the most 
interesting cases for studying the complexity of stellar populations in dwarf 
galaxies. 
%
This galaxy was one of the first dSph in which an intermediate age
population was detected. The presence of upper asymptotic giant branch
(AGB) stars, brighter and redder than the tip of the red giant branch
(RGB), indicated that about $20\%$\ of the galaxy could be of
intermediate age (2 to 8~Gyr) (Aaronson \& Mould \cite{aaro+moul80},
\cite{aaro+moul85}).  Surveys for AGB stars led to discovery of 111
carbon stars, for which follow up near-infrared photometry indicated a
wide range of bolometric luminosities, a mass dispersion among the
progenitors, and hence an age spread (Frogel et al.
\cite{frog+82}; Westerlund et al. \cite{wester87}; Lundgren  \cite{lundgren}; 
Azzopardi et al. \cite{azzo+99}).
Fornax is also known to contain a planetary nebula whose abundance patterns are 
consistent with an origin from a second or third generation star (Danziger et 
al. 
\cite{danz78}; Maran et al. \cite{maran}).
The presence of such an intermediate-age population is confirmed by a
conspicuous red HB clump (Demers et al. \cite{deme+94}; Stetson et
al. \cite{s98}).  
Most recently, an HST study of a central Fornax field sampling the
main-sequence turnoffs of the intermediate-age and old populations has
been carried out by Buonanno et al. (\cite{buon+99}). The analysis of the
resulting CMD has shown evidence for a star formation starting about
12 Gyr ago and continuing until 0.5 Gyr ago.
A variable star formation rate is revealed by gaps 
between separate subgiant branches, and major star formation episodes probably 
occurred nearly 2.5, 4, and 7 Gyr ago. 

Also, Fornax certainly harbors an old stellar population, since it
contains five globular clusters whose ages do not differ from those of
M68 and M92 (Buonanno et al. \cite{buon+98}; Smith \etal
\cite{smit+98}), except perhaps for cluster~4 that appears to be 2-3
Gyr younger (Buonanno et al. \cite{buon+99}).  These clusters have
unusually red horizontal branches for their low metallicity, with no
counterparts in the outer Galactic halo or the Magellanic Clouds.
Also for cluster 4, the recent WFPC2 color-magnitude diagrams of
Buonanno et al. (\cite{buon+99}) unambiguously indicate a low
metallicity, [Fe/H]$\approx-2$, although integrated spectra pointed to
a metallicity similar to that of field stars (Beauchamp et
al. \cite{beauchamp}).  An old population is present among the field
stars of Fornax as well, as demonstrated by detection of a red
horizontal branch slightly fainter than the red clump, and of RR Lyrae
variables (Buonanno et al. \cite{b85}; Stetson et al. \cite{s98},
hereafter \abbrev{SHS98}).

Fornax also hosts a significant population of young stars. Buonanno et
al.  (\cite{b85}) had already noticed a handful of faint blue stars in
their \cmd\ of the Fornax field, tentatively explained as belonging to
a $\sim 2\times10^9$~yr population (\cf also Gratton \etal
\cite{grat+86}). The deeper \cmd\ of Beauchamp \etal
(\cite{beauchamp}) clearly revealed a young main-sequence, and
comparison with theoretical isochrones indicated recent star
formation. The brightest turnoff was located at $M_V \simeq -1.4$,
implying a minimum age of $\sim 10^8$~yr. This young population is
best shown by the recent wide-area survey of 
\abbrev{SHS98}.
Notwithstanding this young stellar component, Fornax appears to be
devoid of any interstellar medium (ISM). A large-area search for neutral
hydrogen has given no detectable
\hi emission or absorption (Young \cite{youn99}), the upper limit for \hi 
emission 
being $5 \times 10^{18}$\ cm$^{-2}$\ at the galaxy center.  Thus the
interstellar medium that must have been present a few $10^8$\ yr ago
to form stars, appears to be all gone. 
There is also the possibility that the ISM has been ionized and heated
up by the interstellar UV field. However, this hypothesis conflicts
with the lack of detection of X-ray emission in the direction of
Fornax (Gizis \etal \cite{gizi+93}).

The various stellar subpopulations in Fornax have different spatial
distributions, which have been carefully investigated by 
\abbrev{SHS98}.  The oldest population, represented by the RR~Lyrae
variables, has the most extended distribution. The intermediate-age
stars (red clump stars) are more centrally concentrated, while the
young population of blue MS stars, as well the reddest AGB (carbon)
stars, are even more concentrated in a bar-like distribution roughly
aligned in the EW direction, with the brightest stars located at the
ends of the bar.  
Also the red clump population displays an
asymmetrical structure (\cf Hodge
\cite{hodge61b}; Eskridge \cite{eskridge2}; Demers et
al. \cite{deme+94}), with a peculiar ``crescent'' shape (SHS98).

Despite all these pieces of knowledge accumulated in recent years, the
star formation history of Fornax is not yet fully understood. Several
questions need to be answered before a reconstruction of the star
formation and chemical enrichment history of Fornax can be
attempted. The metallicity should be measured for stellar populations
of different age and location within the galaxy, so as to determine
the run of metal enrichment as a function of time. The star formation
history needs to be evaluated using critical features in the \cmd\ as
tracers of star formation at different epochs, to understand to what
extent star formation proceeded continuously or in bursts, and how it
propagated throughout the body of the galaxy. The nature of the wide
red giant branch (RGB) is still quite puzzling, though all previous
investigations agree on the fact that it is broader than expected on
the basis of the photometric errors. Further, there is a lack of
observational data with which to study features such as the RGB and
AGB bumps or the precise location of central helium-burning stars as a
function of age and metal abundance, as a test of stellar evolution
models. Large field observations of Local Group dSph galaxies, being
able to sample a significant number of stars, can address these
issues.

With these open questions in mind, we have investigated the stellar
populations of Fornax as part of a wide-field study of nearby dwarf
spheroidals. We present here a large area $BVI$\ photometric study of
the Fornax field, yielding magnitudes and colors for about 40,000
stars down to $\sim2$\ mag below the horizontal branch, in four
regions located at different distances from the galaxy center.  The
use of standard passbands, together with the size of our stellar
sample, allowed us to derive the basic physical properties of Fornax
with high accuracy and measure details in its \cmd\ that trace the
less numerous populations and faster evolutionary phases.

In particular, the $B$\ band turned out to be invaluable for studying
the hot stars, be they old or young, whereas the wide baseline of the
$(B-I)$\ color provides the best resolution of the different
evolutionary phases in the color-magnitude diagrams (\cf Smecker-Hane
\etal \cite{smec+94}; Held et al. \cite{held+99}).  Also, the
availability of a comparison field allowed us to estimate the
foreground and background contamination.
The present photometry will be the input database to model the star
formation history (SFH) of Fornax (Held et al., in preparation) using
population synthesis techniques.

The paper is organized as follows. The observations and data reduction
are presented in Sect.~\ref{sec_obsred}.
In Sect.~\ref{sec_cmd} we present $V$, $(B-I)$\ color-magnitude
diagrams of the Fornax field stars and discuss several interesting
features with the help of theoretical isochrone fitting.
The $V$, $I$\ luminosity function is derived in Sect.~\ref{sec_lf} and
used to estimate the distance to Fornax. This is confirmed by an
independent distance estimate based on the $V$\ luminosity of old-HB
stars (Sect.\ref{sec_disthb}).
In Sect.~\ref{sec_metal} we compare the \cmd\ of Fornax with template
globular cluster RGB sequences using the standard $(V-I)$\ colors, and
discuss the mean abundance and age of the dominant population.
The color distribution of red giant stars in Fornax is analyzed in detail in 
Sect.~\ref{sec_coldistrib}, where the size of an intrinsic abundance spread is 
discussed. 
Some light on the nature of the wide RGB of Fornax is shed by a comparison of 
the 
spatial gradient of different populations (Sect.~\ref{sec_popratios}). 
Our results and conclusions are summarized in Sect.~\ref{sec_conclu}. 

\section{Observations and data reduction}
\label{sec_obsred}

 \tablecaption{The journal of observations \label{t_obsjou}}
 \begin{planotable}{llcrcc}
 \tablehead{
 \colhead{Nt.} &
 \colhead{ID}  &
 \colhead{Filter} &
 \colhead{$t_{\rm exp}[{\rm s}]$} &
 \colhead{$X$} &
 \colhead{FWHM$[\arcsec]$} }
 \startdata
  19 Oct. 1995 &  A       &  $B$  & 3$\times$900 & 1.02 & 1.5     \\ %
  19 Oct. 1995 &  A      &  $I$  & 3$\times$900 & 1.11 & 1.5     \\ %
  19 Oct. 1995 &  A     &  $V$ & 3$\times$900 & 1.06 & 1.7    \\ %
  20 Oct. 1995 &  B      &  $B$ & 3$\times$1200 & 1.09 & 1.5  \\ %
  20 Oct. 1995 &  B      &  $I$ & 1200 & 1.01 & 1.3         \\ %
  19 Oct. 1995 &  B      &  $V$  &    600 & 1.26 & 1.4   \\ %
  20 Oct. 1995 &  B      &  $V$  & 2$\times$600 & 1.22 & 1.4   \\ %
  21 Oct. 1995 &  C    &  $B$  & 3$\times$1200 & 1.25 & 1.5    \\ %
  21 Oct. 1995 &  C    &  $I$ & 3$\times$1200 & 1.03 & 1.3        \\ %
  20 Oct. 1995 &  C    &  $I$ &      240 & 1.11 & 1.0         \\ %
  21 Oct. 1995 &  C    &  $V$ & 3$\times$600 & 1.10 & 1.6        \\ %
  20 Oct. 1995 &  D   &  $B$ & 1200 & 1.34 & 1.5          \\ %
  21 Oct. 1995 &  D   &  $B$ & 2$\times$1200 & 1.03 & 1.5       \\ %
  21 Oct. 1995 &  D   &  $I$ & 3$\times$1200 & 1.01 & 1.3        \\ %
  21 Oct. 1995 &  D   &  $V$ & 3$\times$600 & 1.11 & 1.6         \\ %
  21 Oct. 1995 &  BKG  &  $B$ & 1800 & 1.15 & 1.6    \\ 
  21 Oct. 1995 &  BKG  &  $I$  & 1500 & 1.24 & 1.4    \\ 
  21 Oct. 1995 &  BKG  &  $V$ &  900 & 1.36 & 1.8     \\ 
 \end{planotable}

 \begin{figure}[t]
 \vspace{0cm}
 \hbox{\hspace{0cm}\psfig{figure=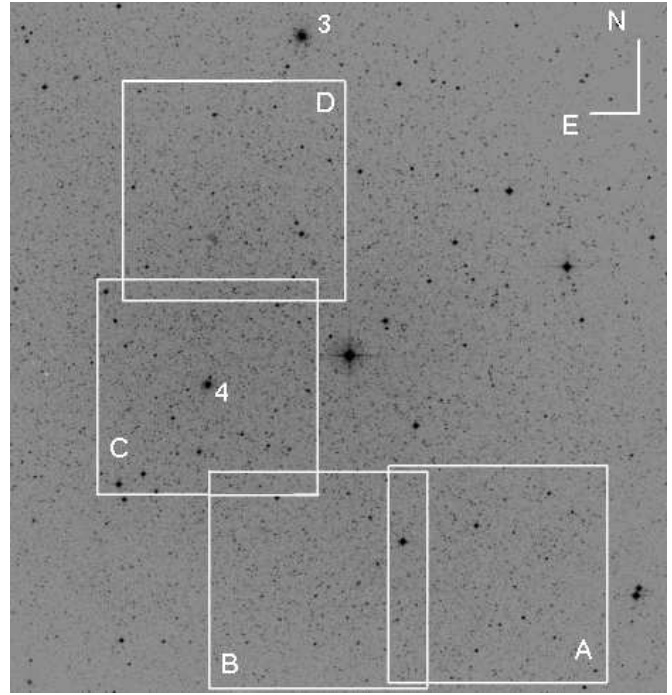,width=8.8cm,bbllx=70pt,bblly=150pt,bburx=540pt,bbury=640pt} }
 \vspace{0cm}
 \caption[]{The central area of Fornax reproduced from the Palomar Digital Sky 
 Survey. The squares indicate the $10\farcm7\times10\farcm7$\ regions
 studied in this paper. The Fornax globular clusters \#3 and \#4 are also
 indicated}
 \label{magmap}
 \end{figure}

\subsection{Observations}
The Fornax galaxy was observed on October 19--21, 1995 using DFOSC at
the ESO/Danish~1.54m telescope. The detector was a $2048\times2048$
Loral CCD with pixel 0\farcs40, covering a field of view of $13\farcm
6 \times 13 \farcm 6$.  Due to non-uniform sensitivity near the edges
of the CCD, the images were trimmed to a useable area of $1600\times
1600$~pixels (i.e. $10\farcm 7 \times 10\farcm 7$).
%
CCD readout by amplifier B in high--gain mode yielded a noise of
7.2~e$^{-}$/px (rms) and a conversion factor of 1.31~e$^{-}$/ADU.

We observed 4 slightly overlapping fields in Fornax, plus one control
field. A map of the location of the fields is shown in
Fig.~\ref{magmap}. The innermost field (region C) is centered on the
globular cluster \#4, \ie\ at about 4 arcmin from the galaxy center
(as defined by \abbrev{SHS98}). The outermost field (A) is located at
$\sim13\farcm5$\ from the center.
The journal of the observations is reported in Table~\ref{t_obsjou}.
The columns give the night, an image identifier, the filter, exposure
time and airmass, and the FWHM of the point spread function (PSF). The
seeing was only fair, yet adequate to measure the relatively bright
stars in our database.  The third night had the most stable weather
conditions. Several short exposure images, not included in this table,
were used for checking the photometric zero points.

\subsection{Reduction and photometry}
The image processing was carried out with the {\sc eso/midas} package
in a standard way. Reduction follows the procedures detailed by
Saviane et al.  (\cite{savi+96}, Paper~I) and Held et
al. (\cite{held+99}, Paper~II).
For each field/filter combination, master images were produced by
registering and coadding the long exposure images. The PSF was not
significantly degraded by this process.
Stellar photometry was performed using {\sc daophot} and {\sc allstar}
(Stetson \cite{stetson87}). 
%
%
The final PSF star catalogs contained $\simeq 50$~stars, and the best
fit was obtained by fitting a Moffat ($\beta = 1.5$) function with a
quadratic dependence on the $x$, $y$ star coordinates.
%
{\sc allstar} was run twice on the sum images. In the second run, 
the star subtracted frames were searched for faint undetected objects that were 
added 
to the input lists of stars. The master photometric catalogs were
created using the complete lists of stars as new inputs to {\sc allstar}. 
%


\subsection{Calibration}

Observations of Landolt's (\cite{landolt92b}) standard star fields were 
used to calibrate the photometry. 
The raw 
magnitudes were first normalized according to the following equation
\begin{equation}\label{eqnorm}
m' = m_{\rm ap} + 2.5\, \log (t_{\rm exp} + \Delta\,t) -
k_{\lambda}\,X
\end{equation}
where $m_{\rm ap}$ are the instrumental magnitudes measured in a
circular aperture of radius $R = 6\farcs9$, $\Delta\,t$ is the shutter
delay and $X$ the airmass. 
A shutter delay  of $-0.11$~s  was estimated from a sequence of images 
taken with increasing exposure times.
The extinction coefficients $k_B = 0.235$, $k_V = 0.135$ and $k_I = 0.048$\ were 
adopted from the Geneva Observatory Photometric Group data. 
%
The normalized instrumental magnitudes were then compared to
Landolt's (\cite{landolt92b}) values, and the following relations were
found:
\begin{eqnarray}
B &=&  b' + 0.207 \, (B-V) + a_B     \\
V &=&  v' + 0.0544 \, (B-V) + a_V   \\
V &=&  v' + 0.0489 \, (V-I) + c_V    \\
I  &=&   i' - 0.00658 \, (V-I) + a_I 
\end{eqnarray}
where $a_B = 23.041$, $23.025$ and $23.038$ for the nights 1, 2 and 3,
respectively. In the same order of nights, the other coefficients are
$a_V = 23.774$, $23.763$ and $23.772$; $c_V = 23.772$, $23.762$ and
$23.771$; and finally $a_I = 23.070$, $23.054$ and $23.077$. The
standard deviations of the residuals were 0.018, 0.013, and 0.022 mag
in $B$, $V$ (both equations), and $I$\ respectively.

The PSF magnitudes were scaled to aperture magnitudes by assuming that
$m_{\rm ap} = m_{\rm PSF} + const.$ (Stetson \cite{stetson87}).
Aperture magnitudes were measured for  a sample of bright, isolated
stars, and for each star we computed the difference with respect to
the PSF magnitude measured on the coadded frames. The same reference
aperture used for the standard stars was employed. The internal
calibration uncertainty due to ``aperture correction'', estimated from
the consistency of the zero point derived from several individual
images, is of the order 0.01 mag for all filters.

The instrumental magnitudes and colors for the stars observed in at
least two filters were calibrated either with an iterative procedure
or by solving a system of 3 equations of the form
\[
m_{\rm std} = m_{\rm inst} + k_m \, color_{\rm std} + a_m
\]
where $m_{\rm std}$ and $color_{\rm std}$ are the magnitude and color
in the standard system,  and $m_{\rm inst}$ are the instrumental
magnitudes. 
%

After independent calibration, we performed a verification of the
photometric zero points of the catalogs in the 4 zones by comparing
stars in the overlap strips. Since the mean systematic deviations
between $BVI$\ magnitudes measured in the field C and D (both observed
during the third, most stable night) are less than 0.03 mag, we chose
to refer all photometry to the zero point of the central field C, and
applied small zero-point corrections to our photometry in the field A,
B, and D. We conservatively adopt an uncertainty of 0.03 mag as our
systematic error in all bands.

\subsection{Comparison with previous studies}

As a further check of the accuracy of our photometric zero point, we
compared our results with previous data in the literature. The only
published photometry tables are those of Buonanno et
al. (\cite{b85}). Their Tables 6 and 7 report the values of $V$ and
$(B-V)$ for all the stars measured in two separate
$2\times2$~arcmin$^2$ fields, called A1 and A2, which are included in
our fields C and A.
The two sets of measurements for the A2-A field pair are in good agreement.  
The median differences (this paper -- Buonanno et al.) are 
$ -0.013$ in $V$
and $ -0.017$ in $B-V$, with standard deviations of 0.16 mag
in both cases.


The consistency between the zero-points for the A1-C pair is
still good, yielding median residuals 0.014 ($\sigma=$0.20) in $V$
and 0.024 (0.25) in $B-V$.

\subsection{Artificial star tests}

 \tablecaption{
 The photometric errors from artificial star experiments 
 \label{t_err}}
 \begin{planotable}{rrrr|rrr}
 \tablehead{
 \colhead{$V,B,I$} &
 \colhead{$\sigma_V$} &
 \colhead{$\sigma_B$} &
 \colhead{$\sigma_I$} &
 \colhead{$\sigma_V$} &
 \colhead{$\sigma_B$} &
 \colhead{$\sigma_I$} 
 }
 \startdata
 &       &    A  &        &       &    C  &        \\
  14.75 &    \nodata &    \nodata & 0.003  &    \nodata & \nodata &  \nodata  \\ 
  15.25 &    \nodata &    \nodata & 0.003  &    \nodata &  \nodata & \nodata  \\ 
  15.75 &    \nodata &    \nodata & 0.003  &    \nodata &    \nodata & 0.011  \\ 
  16.25 &    \nodata &    \nodata & 0.004  &    \nodata &    \nodata & 0.012  \\ 
  16.75 &    \nodata &    \nodata & 0.005  &    \nodata &    \nodata & 0.014  \\ 
  17.25 &    \nodata & 0.013 & 0.006  &    \nodata & 0.010 & 0.017  \\ 
  17.75 &    \nodata & 0.014 & 0.013  & 0.014 & 0.014 & 0.018  \\ 
  18.25 & 0.013 & 0.014 & 0.017  & 0.016 & 0.019 & 0.019  \\ 
  18.75 & 0.017 & 0.019 & 0.020  & 0.020 & 0.018 & 0.026  \\ 
  19.25 & 0.021 & 0.024 & 0.028  & 0.025 & 0.023 & 0.032  \\ 
  19.75 & 0.024 & 0.031 & 0.037  & 0.032 & 0.030 & 0.045  \\ 
  20.25 & 0.030 & 0.038 & 0.049  & 0.040 & 0.036 & 0.064  \\ 
  20.75 & 0.038 & 0.052 & 0.069  & 0.061 & 0.049 & 0.082  \\ 
  21.25 & 0.057 & 0.069 & 0.098  & 0.081 & 0.064 & 0.108  \\ 
  21.75 & 0.073 & 0.085 & 0.144  & 0.103 & 0.082 & 0.156  \\ 
  22.25 & 0.107 & 0.105 & 0.150  & 0.136 & 0.106 &    \nodata  \\ 
  22.75 & 0.116 & 0.134 & 0.181  &    \nodata & 0.124 &    \nodata  \\ 
 \end{planotable}

 \begin{figure*}[t]
 \resizebox{12cm}{!}{\includegraphics{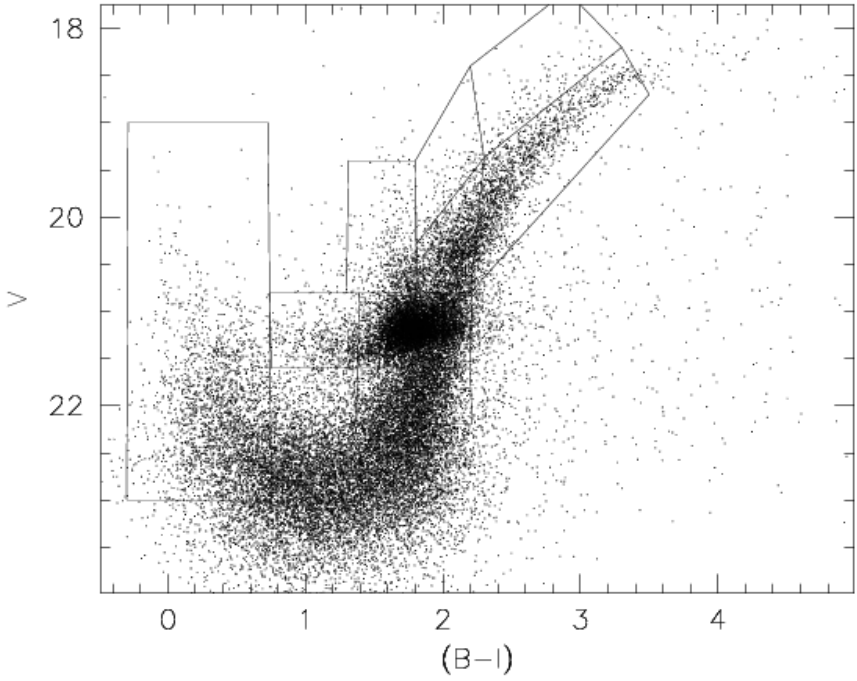} }
 \hfill
 \parbox[b]{55mm}{
 \caption[]{
 The color-magnitude diagram of Fornax in the $(B-I), V$ plane. This diagram 
 includes about 42500 stars in all fields. 
 The most noticeable features are the wide RGB made of old and intermediate-age 
 stars, the upper AGB tail, a young main sequence, and a prominent red clump 
 together with fainter, older HB stars. The blue main sequence clearly merges 
 into a 
 mix of subgiant branches. 
 The outlined regions have been used for counting stars in different evolutionary 
 phases ({\it see text})
 } 
 \label{1cmdBVI}  
 } 
 \end{figure*}

Extensive artificial star simulations were performed to evaluate the
uncertainties of our photometry and the completeness of the data. The
simulations were carried out for the fields A and C, which represent
the lowest and highest crowding in our frames.  A list of input stars
was created for each $V$\ master image, with uniformly distributed
magnitudes.  The star coordinates were generated over a grid of
triangles with a small random offset from the vertices, a
configuration allowing to add the largest number of non-overlapping
simulated stars. Error estimates based on randomly placed artificial
stars may not be realistic if there is a significant amount of
clustering among real stars.  
This caveat does not seem to apply to our relatively uniform 
Fornax fields, though.


The same artificial stars were used in all bands, using random $(B-V)$
and $(V- I)$ colors so that the stars were uniformly distributed in
the color-magnitude diagrams.  We typically added $\sim36000$\ stars
per filter in 80 experiments.
The frames with the artificial stars were then reduced using exactly
the same procedures as for the original images.  For each filter, the
retrieved artificial stars were matched to the input list by means of
their coordinates. The stars recovered in different colors were
matched, and the raw catalog calibrated just as the original
photometry.
%
%
The standard deviations of the measurement errors $\Delta m$,
calculated in 0.5 mag bins, are given in Table~\ref{t_err}. The first
column gives the bin centers, columns 2 to 4 and 5 to 7 list the
errors obtained for the fields A and C, respectively.  The measured
standard errors span a range from $\simeq 0.01$~mag near the tip of
the RGB to $\simeq 0.15$~mag close to the limiting magnitudes.  Errors
and the limiting magnitudes are consistent with the different crowding
conditions and exposure times in the two fields.
%
%
Contour plots of the completeness levels were produced by dividing
each \cmd\ in cells with color and magnitude steps of 
$0.5$ and $0.2$~mag,  
respectively, and counting the artificial stars within each cell before and 
after reduction. We used the same acceptance criteria as for the 
galaxy's \cmd, i.e. a star was counted in a cell only if it was recovered in 
each 
of the 3 filters. The completeness array was computed as the ratio between the 
post-reduction and the input simulated star counts, and median filtered with a 
$5\times5$ box to reduce the noise 
in the contour plots. Examples of the resultant completeness contours will be 
shown 
in Sect.~\ref{sec_cmd}.

\section{Color-magnitude diagrams }
\label{sec_cmd}

 \begin{figure*}[t]
 \resizebox{12cm}{!}{\includegraphics{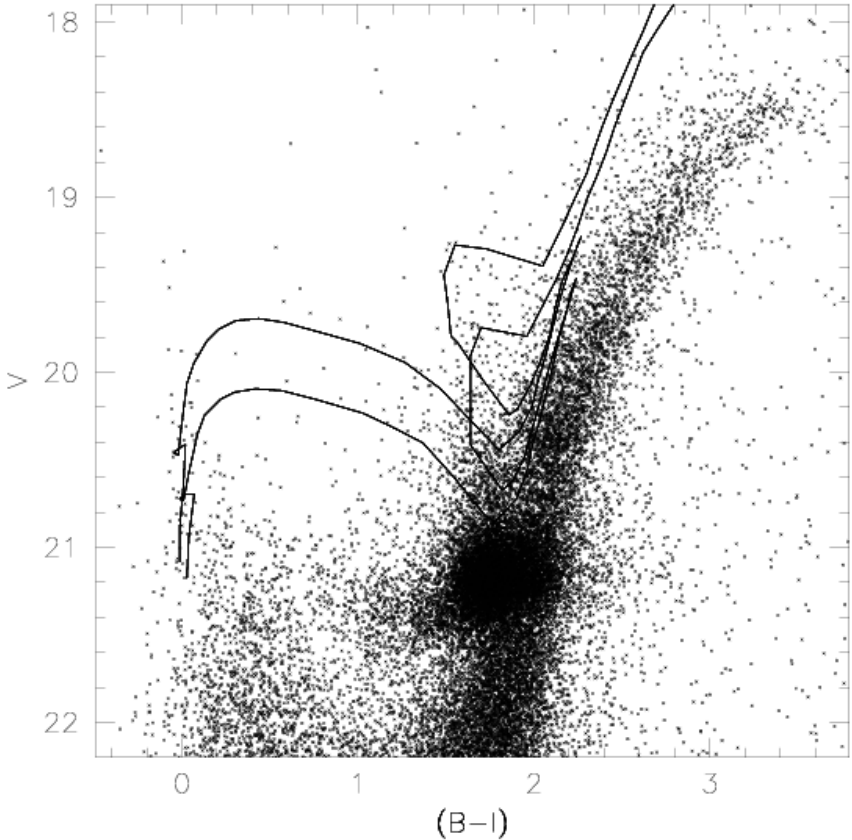} }
 \hfill
 \parbox[b]{55mm}{
 \caption[]{
 An enlarged view of the color-magnitude diagram of Fornax showing the features 
 produced by core helium-burning stars, and morphological details of the red 
 giant 
 branch. Two isochrones from the post-MS models of Bertelli et al. 
 (\cite{bertelli}) 
 with Z=0.004 and ages 300 and 400 Myr (top to bottom) have been superimposed to 
 the data. The plume of stars above the red clump is composed of intermediate 
 mass 
 stars burning helium in the core after leaving the young main sequence.  
 }
 \label{f_bertelli} 
 }
 \end{figure*}


The combination of a wide photometric baseline and large sample size employed in 
this study allowed us a very detailed view of the evolved stellar populations in 
Fornax. Fig.~\ref{1cmdBVI} presents the $V$, $(B-I)$\ color-magnitude diagram 
of this nearby dwarf spheroidal, showing an excellent separation of stars in 
different 
evolutionary phases. 
%
We now briefly describe the many interesting features seen in this \cmd.  

\begin{itemize}
\item{A wide red giants branch, comprising stars older than $\sim1$\ Gyr. The 
color spread is much larger than expected from photometric errors
(cf. Sect.~\ref{sec_coldistrib}). While the red side of the RGB shows
a sharp edge, the stars are spread on the blue side forming sparsely
populated sequences distinct from the RGB (\cf \abbrev{SHS98}).  This
is more evident in Fig.~\ref{f_bertelli}. Note that the foreground and
background contamination is virtually negligible in the relevant
regions of the diagram. The color-magnitude diagram of foreground and
background objects in our control field, covering an area $10\farcm7
\times 10\farcm7$, contains only 154 objects down to $V\approx22.5$.
}
\item{Above the RGB tip at $V \sim 18.4$, there is a well-developed upper AGB 
tail extending to colors as red as $(B-V)$ and $(V-I) \sim 3$, or
$(B-I)
\sim 6$. The upper AGB consists of intermediate-age C and M stars, the
latter comprising a small group of stars just above the RGB tip (see
SHS98).  }
\item{A rich red clump contains the majority of HB stars of a numerous 
intermediate-age and metal-enriched population; in the following we
will refer to it simply as red clump (RC) (\cf Demers et
al. \cite{deme+94};
\abbrev{SHS98}). 
About 0.2 mag fainter, a horizontal branch originating from an older
population is clearly seen, indicated in the following as ``old HB''
(or \oldhb). We also notice the instability strip mostly populated by
RR~Lyrae variables, whose random phase colors and magnitudes define a
band $\lesssim 1$\ mag thick. RR Lyrae variables are present in the
Fornax field (Buonanno et al. \cite{b85}; SHS98; Mateo
\cite{mate98}). Blue HB stars are hardly seen in this diagram. If any
exist, they are confused with the young main sequence stars.  }
\item{A blue plume reaching $V\sim20$, identified with a young main sequence of 
$\sim0.1$\ Gyr old stars (Beauchamp \etal \cite{beauchamp};
\abbrev{SHS98}).  }
\item{An almost vertical plume originating from $(B-I)\sim1.8$, $V\sim21$, i.e. 
just above the red clump, extending up to $V\sim19.3$. They are, as it
will be shown below, core helium-burning stars with mass in the range
$\sim2$\ \msol.}
\item{At fainter magnitudes, the main sequence merges into a heavily populated 
region $\sim 1$~mag below the HB, involving a mix of subgiant branches
of different ages.  }
\item{
We also note the small clump of stars at $V\sim20.4$\
on the red giant branch, an example of the short-lived evolutionary
phases that can be revealed in stellar populations using adequately
large data samples. This feature is identified with the AGB bump, a
clumping of stars due to a slowing down of the luminosity increase at
the beginning of AGB evolution (\eg Gallart
\cite{gall98}; 
Alves \& Sarajedini \cite{alve+sara99}). We will return on this point in 
Sect.~\ref{sec_lf}.
}
\end{itemize}
Overall, this diagram shows that Fornax went on forming stars from an early 
epoch 
($>10$\ Gyr ago) until about 100 Myr ago. 
%
A closer picture of the \cmd\ is presented in Fig.~\ref{f_bertelli},
showing details of the core helium-burning stars, which are important
diagnostics of stars formation histories in galaxies. The prominent
red clump comprises stars with age 2--10 Gyr and mass approximately
0.9 to 1.4 \msol. Its mean luminosity and color bears information on
the mean age of Fornax. We will return on this point later. A
comparison with theoretical isochrones of Bertelli et
al. (\cite{bertelli}) allows one to establish the nature of the stars
producing the plume above the RC. These are intermediate-mass core
helium-burning stars (2.4--2.9 \msol), counterparts of the young main
sequence stars in the age range 0.3--0.5 Gyr, which started burning
helium in a non-degenerate core.

This stage (also known as the ``blue-loops'') represents an important
indicator of the metallicity of the young population and therefore of
the chemical enrichment history of galaxies (\eg Aparicio et
al. \cite{apar+96}; Cole et al \cite{cole+99}; and references
therein).
Fig.~\ref{f_bertelli} shows that fitting the RC plume requires isochrones 
having 
[Fe/H]$\sim-0.7$, i.e. significantly more metal-rich than the bulk of the Fornax 
stellar population. 
This relatively high metallicity of the younger stars may probably 
explain why Fornax seemingly lacks a large population 
of anomalous Cepheids (\abbrev{AC}), which are so numerous in the 
metal-poor dSph Leo~I (see discussion of the instability strip 
in Caputo et al. \cite{capu+99}). Searches of AC's in Fornax are 
underway (Bersier \& Wood \cite{bers+wood99}). 

 \begin{figure}[t]
 \vspace{0cm}
 \psfig{figure=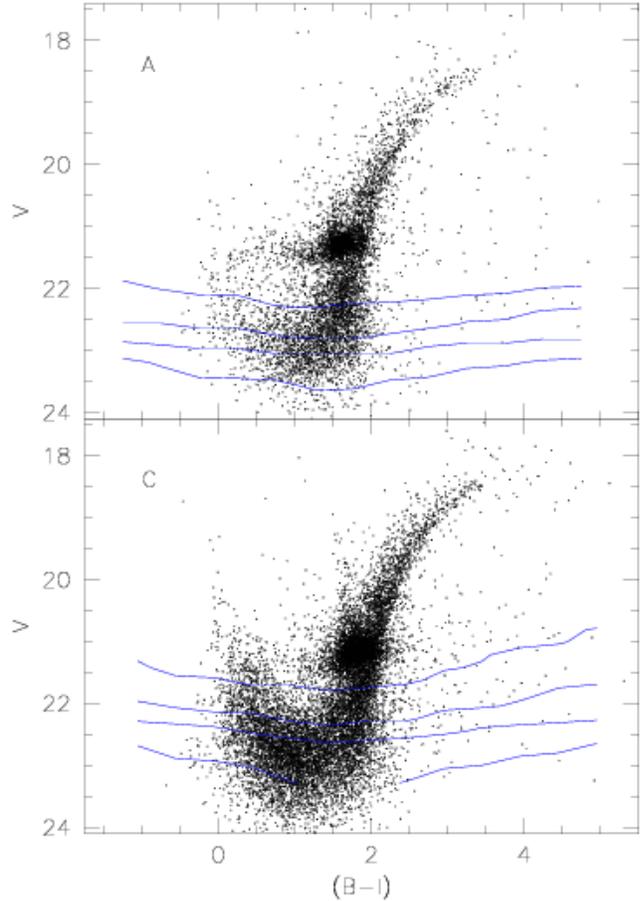,width=8.8cm}
 \vspace{0cm}
 \caption[]{A comparison of the color-magnitude diagrams in an outer and inner 
 region of Fornax. Field A (top panel) is located about 14\arcmin\ from
 the center of Fornax, while field C (bottom) samples an area near the
 galaxy center.  The contour lines represent the 30\%, 50\%, 70\% and
 90\% completeness levels of our photometry. Note that the completeness
 in the crowded field C is lower than in field A. The number of young
 main sequence stars and their helium-burning counterparts decreases
 from the inner to the outer field, allowing perceiving the bluer HB
 stars belonging to the oldest Fornax field population }
 \label{f_cmdAC} 
 \end{figure}


The color-magnitude diagrams for the innermost and outermost region in this 
study,  
shown in Fig.~\ref{f_cmdAC}, illustrate the remarkable variation of the stellar 
populations in Fornax with galactocentric radius. 
%
%
The inner field (C, bottom panel) shows all of the features noticed in the total 
\cmd. 
The young main-sequence stars are less numerous in the outer field,
even accounting for the lower stellar surface density, and there are very few 
blue 
stars brighter than $V\sim22$. The young core-He-burning stars (the plume above 
the RC) follow the trend of the blue main sequence stars. 
In contrast, it is interesting to note a hint of a {\em blue horizontal branch} 
in the 
outer field. Together with the detection of RR Lyrae stars, our data provide 
evidence 
for a small {\it old, metal-poor field population} similar to that of the Fornax 
globular clusters. 
A population~II halo seems to be common not only in dwarf spheroidals (see Mateo 
\cite{mate98}) but also in dwarf irregulars (\eg Minniti et al. \cite{minn+99}; 
Aparicio et al. \cite{apar+97}). 
A quantitative estimate of the population gradient in Fornax will be given in 
Sect.~\ref{sec_popratios}.

\section{Analysis and discussion}
\label{sec_discu} 

\subsection{Luminosity function and distance}
\label{sec_lf}

 \begin{figure}[t]
 \psfig{figure=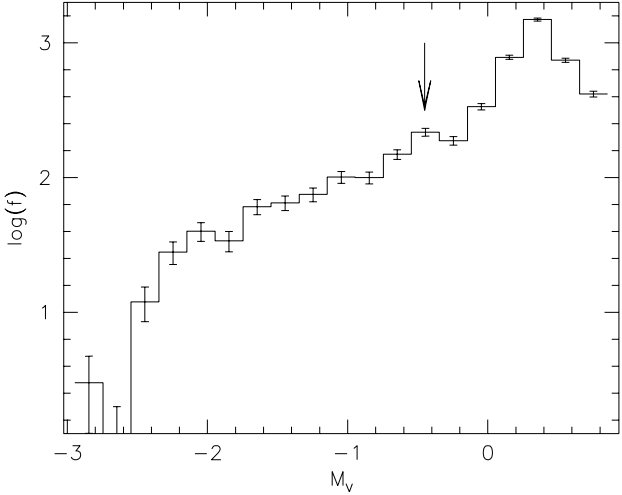,width=8.8cm}
 \caption[]{
 The $V$\ RGB luminosity function of Fornax, showing the sharp cutoff at the RGB 
 tip, the prominent red clump at $M_V \approx +0.4$, and the ``AGB bump'', a 
 clumping of intermediate age stars at the beginning of their AGB double-shell 
 burning phase ($M_V \approx -0.4$) 
 }
 \label{f_lf}
 \end{figure}

The red giant luminosity function (LF) and distance to Fornax was derived using 
stars within  $\pm 2\sigma$ from the fiducial sequence. 
As shown in Sect.~\ref{sec_coldistrib}, this implies selecting the dominant 
stellar 
population in Fornax. Luminosity distributions were obtained both in $V$ and in 
$I$ 
by counting stars in 0.2~mag bins down to below the red clump.  
Since at these bright magnitudes our photometry is virtually complete, there was 
no 
need to correct the observed LF's for incompleteness. Foreground and background 
contamination is not a concern, either, because the number of field objects in 
the 
proximity of the RGB is negligible. 

The cutoff in the $I$\ luminosity function corresponding to the maximum 
luminosity 
reached by red giants before they ignite the He burning, has proven to be a good 
distance indicator (Madore \& Freedman \cite{madoreFreedman95}; see also Salaris 
\& Cassisi \cite{sala+cass98}). 
We measured the $I$ magnitude of the RGB tip separately for our 4
Fornax fields, following the methods of Lee \etal (\cite{lee_etal93}).
The mean of the four values is $<I_{\rm TRGB}> = 16.72 \pm 0.10$.
Although the scatter of the individual measurements is small
($\sim0.02$\ mag), we have adopted a larger uncertainty to take into
account both the intrinsic precision of the tip detection method,
which is about a half of the 0.2 mag bin, and the zero point
uncertainties.

We then calculated the distance to Fornax using the relations of 
Da Costa \& Armandroff (\cite{daco+arma90}). 
This procedure, often applied to derive the distance of dwarf galaxies, 
implicitly 
assumes that the $I$ magnitude of the tip is little affected by age.
Theoretical models indeed show a dependence of the tip luminosity on the age, 
but 
this is more pronounced for very metal-poor populations ([Fe/H]$<-1.7$) and ages 
younger than 5 Gyr (\eg Caputo et al. \cite{capu+99}).

%
%
The relations of Da Costa \& Armandroff (\cite{daco+arma90}) give the $I$ 
bolometric correction as a function of color of the stars near the RGB tip, and 
the bolometric luminosity of the tip as a function of metallicity. 
The dereddened color of the RGB tip in Fornax, calculated as the median ($V-I$) 
within 0.1 mag from the tip, is $(V-I)_{0, {\rm TRGB}}= 1.59 \pm 0.06$,
where the error reflects the scatter of the values obtained in our four fields 
plus the 
absolute zero point uncertainty.
We adopted a reddening $E_{B-V} = 0.03 \pm 0.03$\ from Burstein \& Heiles 
(\cite{bursteinHeiles82a}),  yielding $E_{V-I} = 0.038 \pm 0.038$\ and 
$A_{I} = 0.058 \pm 0.058$.
%
%
The bolometric correction is then $BC_I = 0.495 \pm 0.015$, 
while a metallicity ${\rm [Fe/H]} = \feh \pm \errfeh$\ (cf. 
Sect.~\ref{sec_metal}) 
implies $M_{\rm bol}^{\rm TRGB} = -3.55 \pm 0.01$.
%
We thus obtain $M_I^{\rm TRGB} = -4.04 \pm 0.02$, 
%
and a distance modulus $(m-M)_0 = 20.70 \pm 0.12$, 
corresponding to $138 \pm 8$\  kpc. 

Previous distance estimates range from
$(m-M)_0=20.59\pm0.22$ (Buonanno et al. \cite{b85}) to 20.76  
(Demers et al. \cite{demers90}; Buonanno et al. \cite{buon+99}). 
Sagar et al. (\cite{sagar}) found $(m-M)_0=20.7$. The present estimate therefore 
confirms earlier results. This value is also consistent with the distance moduli 
of 
Fornax globular clusters (Buonanno et al. \cite{buon+98}), yielding an average 
modulus $(m-M)_0 = 20.62 \pm 0.08$. 

Using this distance estimate, we plot in Fig.~\ref{f_lf} the $V$
luminosity function of the red giant stars in the inner region of
Fornax. Besides the obvious red clump at $M_V \simeq 0.4$, we notice
the small yet significant peak near $V=20.4$\ that we identify with
the AGB bump, a clustering of stars that occurs at the beginning of
helium shell-burning evolution. Gallart (\cite{gall98}) has recently
discussed the presence of this feature in the LMC and M\,31 where its
location agrees with the prediction of stellar evolution models
(Bertelli et al. \cite{bertelli}).
We have measured the location of the AGB bump in Fornax by performing a 
Gaussian fit to the LF in the region of the bump. We found  $V=20.40\pm0.04$, 
where the error is mainly set by the zero point uncertainty. This corresponds to 
a 
luminosity $M_V \simeq -0.39\pm0.04$, with an additional 0.1 mag uncertainty on 
the $A_V$\ extinction. Similarly, we measured a mean $V$\ magnitude for the red 
clump 
$V_{\rm RC}=21.18\pm0.04$\ mag, corresponding to $M_V=+0.39$.
Thus the detected clump is $0.78\pm0.06$\ mag brighter in $V$\ than the red HB 
clump. For an assumed age of 5 Gyr and the mean metallicity of Fornax, these 
measurements confirm the identification with the AGB bump and rule out 
alternative 
identifications with the RGB bump. 
The RGB bump is expected to be near the HB level for a metallicity 
[Fe/H]$\approx-1$\ and age 5 Gyr (Alves \& Sarajedini \cite{alve+sara99}).
In Lyndsay~113, a 5 Gyr old cluster in the Small Magellanic Cloud having 
metallicity comparable with that of Fornax, the RGB bump is found to be 
$\sim0.15$\ mag brighter than HB stars in the same cluster
(Mighell et al. \cite{migh+98}). 
Clearly, stars in the RGB bump will be outnumbered by the overwhelming red 
clump. 
Observational data like those presented here for the Fornax dwarf are
important to constrain evolutionary models, which in turn are
necessary to interpret the stellar population of Local Group galaxies.

\subsection{Distance based on the old horizontal branch}
\label{sec_disthb}

An independent estimate of the distance to Fornax was obtained from
the mean level of its old-\abbrev{HB} field stars.  The mean magnitude
of the HB was measured by fitting a Gaussian to the $V$\ mag
distribution of the stars in the range $21.2<V<21.7$, $1.1<B-I<1.5$.
%
We necessarily included only the red part of the \oldhb, since the 
bluer horizontal-branch 
stars appear to be mixed with the blue stars on the young main
sequence.  Note, however, that we do not include any RC stars
(which are brighter than the \oldhb\  and RR Lyrae variables). The mean
level of the red HB is $V_{\rm HB}= 21.37\pm0.04$, where the
uncertainty reflects the scatter of the values measured in the
different fields (larger than the formal error on the mean magnitude),
and the systematic error of the $V$\ zero point. Buonanno \etal
(\cite{buon+98}) found a mean level $V_{\rm HB}=21.25\pm0.05$\ for the
 \oldhb of four globular clusters in Fornax.

Using this value for $V_{\rm HB}$, and $A_V = 3.2~E(B-V)=0.096$, we
calculated the distance modulus of Fornax on the Lee \etal
(\cite{ywlee+90}) distance scale, using their relation for the
absolute visual magnitude of RR Lyrae variables,
\begin{equation}
M^{\rm RR}_V=0.17~ {\rm [Fe/H]} + 0.82
\end{equation}
%
\noindent
for a helium abundance of $Y=0.23$. 

Assuming for red HB stars the relatively metal-rich nominal metal
content of RGB stars, [Fe/H]$\approx-1.4$, this relation would give
$M^{\rm RR}_V=0.59$\ mag and a distance modulus $(m-M)_0 =
20.69\pm0.04$\ for a population with age comparable to that of
Galactic globular clusters. This uncertainty includes internal and
photometric errors only.
However, the mean metallicity of the old-HB stars is probably
lower. If the old population in Fornax is relatively metal-poor, of
the order [Fe/H]$\approx-1.8$\ (as we suggest in
Sect.\ref{sec_popratios}), the relation given by Lee \etal
(\cite{ywlee+90}) would imply $M^{\rm RR}_V=0.51$\ mag and a distance
modulus $(m-M)_0 = 20.76\pm0.04$.  The level of the red HB (distinct
from the clump) is probably the result of contributions from stars in
a range of ages and metallicities. For this reason we refrained from
applying any uncertain correction to convert the measured mean
magnitude of the red
\oldhb\  to an equivalent magnitude of RR~Lyrae variables. Further, this
distance modulus based on the HB level is affected by the
uncertainties on the luminosity of HB stars as a function of age and
metallicity. A discussion of the alternative distance scales, however,
is beyond the scope of this paper.

This measurement of the distance to Fornax based on its old horizontal branch 
star luminosity confirms the distance modulus estimated from the RGB tip. This 
consistency is not unexpected, since both the RGB tip method of DA90 and the HB
absolute magnitude obtained for the \oldhb\ are based on the distance scale of 
Lee et al. 
(\cite{ywlee+90}). These two distance measurements use $I$\ and $V$\ magnitudes, 
respectively, which are observationally independent. 

%

\subsection{Mean abundance and age} 
\label{sec_metal} 

 \begin{figure}[t]
 \vspace{0cm}
 \hbox{\hspace{0cm}\psfig{figure=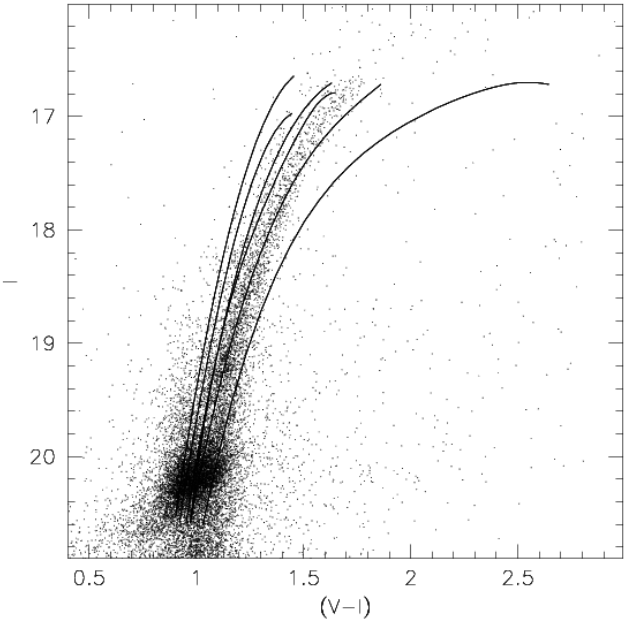,width=8.8cm} }
 \vspace{0cm}
 \caption[]{
 A comparison of our total color-magnitude diagram of Fornax with the giant 
 branches of template Galactic globular clusters from Da Costa \& Armandroff 
 (\cite{daco+arma90}), scaled to the distance and reddening of Fornax. The 
 globular 
 clusters span a metallicity range from [Fe/H]$=-2.2$\ to [Fe/H]$=-0.7$  
 \label{f_da90} }
 \end{figure}

The mean metal abundance of the bulk of the Fornax population was derived by 
direct comparison of the red giant branch in the 
$I$, ($V-I$) color-magnitude diagram with the ridge lines of globular clusters 
from Da Costa \& Armandroff (\cite{daco+arma90})   (see Fig.~\ref{f_da90}).
Our procedure is fully described in Paper~I and II, and is only
briefly outlined here.
In short, we calculated the average color shift, $\delta(V-I)_0$, 
between the Fornax RGB and the Galactic cluster fiducial loci. An interpolation of the relation between the mean color shifts and the globular cluster metallicities (actually a quadratic fit) provides an estimate of [Fe/H] for the dwarf spheroidal. 
This procedure was applied to the $2\sigma$-selected RGB sample (cf.Sect.~\ref{sec_lf}), in two luminosity intervals 
($-4.0 < M_I < -3.0$\ and $-3.0 < M_I < -2.0$), yielding a metallicity 
[Fe/H]$=-1.45 \pm 0.11$\ and [Fe/H]$= -1.33 \pm 0.15$\ dex, respectively. 
%
%
The mean of the abundances determined in these two magnitude bins was adopted as 
our final estimate. The resultant value, 
[Fe/H]$ = \feh \pm \errfeh$, is in good agreement with previous work. We find no 
evidence for a metallicity gradient among the regions studied here, to within 
the errors. 

However, the measurements of mean abundance based on the color of the
RGB are subject to the well-known difficulty in disentangling the effects
of age and metallicity on the effective temperature of red giant stars
(the ``age-metallicity degeneracy'').  Thus we need to estimate the
mean age of Fornax before discussing further its mean metal abundance.
When compared with the predictions of stellar evolution models (\eg
Bertelli et al.
\cite{bertelli}; Caputo et al. \cite{capu+95}), the position 
of core He burning stars in color-magnitude diagrams may provide a useful age
indicator (\eg Caputo et al. \cite{capu+99}; Girardi \cite{gira99};
and references therein). The RC comprises core
helium-burning stars of different ages, so that its location bears
information on the {\it mean age} of the intermediate age population,
weighted by the age distribution function.
%
Thus, similarly to what we had done for the HB, we measured the mean
$(B-I)$\ color in addition to the $V$\ luminosity for the red
clump. The mean magnitude, already reported above, is $V_{\rm
RC}=21.18\pm0.04$\ mag, corresponding to $M_V=+0.39$, in excellent
agreement with Demers et al. \cite{deme+94}).
This means that the RC is $0.19\pm0.06$\ mag more luminous in $V$\
than the old HB stars, a value that appears consistent with the
difference in age of a 13 Gyr old population and a 5 Gyr old bulk
component (see Caputo et al. \cite{capu+99}).
The clump is quite extended in luminosity ($\sim0.6$\ mag), comparable
with that of Carina (Hurley-Keller et al. \cite{hurl+98}), but less
than that of Leo~I (\cf Gallart et al. \cite{gall+99}).
%
The mean color is $<B-I>_{\rm RC} = 1.79\pm0.04$. The uncertainties include the 
field-to-field scatter, comparable with the photometric measurement errors, and 
the zero-point uncertainty. 
The relation  $(V-I) = 0.457~(B-I)+0.147$, obtained from a linear fit to the 
color-color relations for the Fornax red giants in the range $1.0< (B-I) <3.5$, 
yields  $<V-I>_{\rm RC}=0.965$. This value shows excellent agreement with the 
results of Buonanno et al. (\cite{buon+99}). By fitting a parabola to the fiducial points of the RGB, we estimated the interpolated RGB color at the RC level ($V-I \simeq 1.07$\ mag), a value also confirmed by inspection of the WFPC2 color-magnitude diagram (Buonanno et al. \cite{buon+99}).
The difference in color between the red clump and red giant stars at the same luminosity is then 
$\delta_{(V-I),{\rm RC}}=0.10$\ mag, with an estimated uncertainty of 0.02 mag. 
This result can be compared with the model predictions of Girardi
(\cite{gira99}; and priv. comm.) based on the models of Girardi et
al. (\cite{gira+99}), which are in accord with the empirical
calibration of Hatzidimitriou (\cite{hatz91}).  For a metallicity
Z=0.001 (but $\delta_{(V-I),{\rm RC}}$\ is relatively independent of
abundance for metal-poor populations) our result is consistent with a
mean age of the order $5.4\pm1.7$ Gyr. This value is close to the
estimate of Sagar et al.  (\cite{sagar}), based on best fitting of
Yale isochrones, and definitely larger than the age estimated by
Demers et al (\cite{deme+94}).
Most interestingly, the mean age obtained from the clump location appears to be consistent with the presence of MS evolved stars in the same age interval,
as observed with HST (Buonanno et al. 
\cite{buon+99}). This results is quite encouraging for application of this age 
indicator to more distant Local Group galaxies, whose main-sequence turnoff 
cannot be directly measured. 

If we now assume a mean age of approximately 5 Gyr for the bulk of the Fornax 
stars, the observed RGB color would imply a metallicity significantly larger 
than the 
formal result obtained above from a comparison with globular clusters. 
We have estimated the effects of age by comparing theoretical isochrones of 
different 
ages (e.g., 5 and 15 Gyr) at a given metallicity (from Bertelli \etal 
\cite{bertelli}). 
By measuring the ($V-I$) colors at $M_I = -2.5$\ predicted by model
isochrones with Z=0.001 ([Fe/H]$=-1.3$), we find that a 5 Gyr
isochrone is bluer by $\sim0.09$\ mag than a 15 Gyr model
isochrone. This effect mimics a metallicity difference of $\sim 0.4$
dex using the fiducial loci of globular clusters (\cf Paper~II; Caputo
et al. \cite{capu+99}; Gallart et al. \cite{gall+99}).
Thus, if the body of Fornax stars is $\sim5$ Gyr old, the measured
location of the peak of the RGB is necessarily indicative of a higher
mean metallicity, of the order [Fe/H]$=\corfeh$ (clearly the
correction is somewhat model dependent). We regard this value as the
most appropriate estimate of the mean metal abundance of the dominant
stellar population in Fornax. With this correction, the Fornax
metallicity turns out to be very close to that of Sagittarius, a dSph
which has a comparable total luminosity (\eg Bellazzini et
al. \cite{bell+99}).

\subsection{The color distribution of Fornax red giants} 
\label{sec_coldistrib}

 \begin{figure}
 \vspace{0cm}     
 \hspace{0cm}\psfig{figure=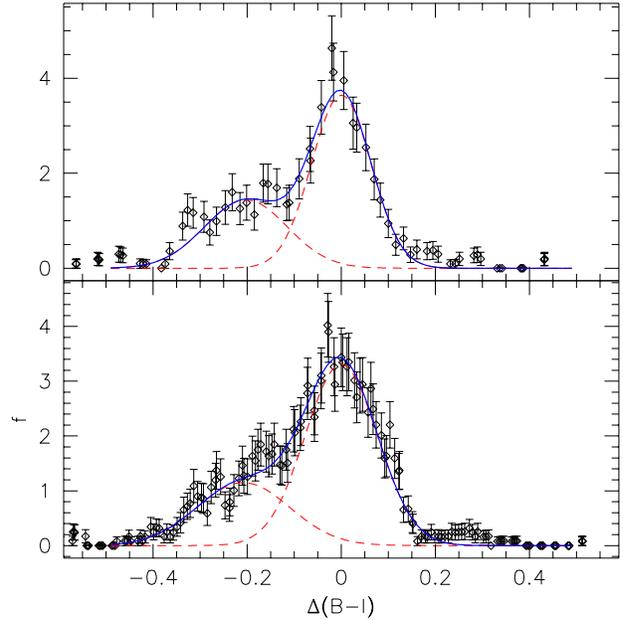,width=8.8cm}
 \vspace{0cm}
 \caption[]{
 The color distribution of the red giant stars in Fornax, plotted
 separately for the inner (bottom panel) and outer region (top
 panel). The histograms represent the distributions of the color
 residuals of individual stars from a median RGB fiducial sequence, in
 the magnitude interval $17.7 \leq I \leq 18.7$.  Error bars represent
 Poisson errors. The color distribution is quite well fitted by the sum
 (continuous line) of two Gaussian functions (dashed lines) suggesting
 a two-component model for the metallicity (or age) distribution of
 Fornax stars.
 }
 \label{twoGauss}
 \end{figure}

 \begin{figure}
 \vspace{0cm}     
 \hspace{0cm}\psfig{figure=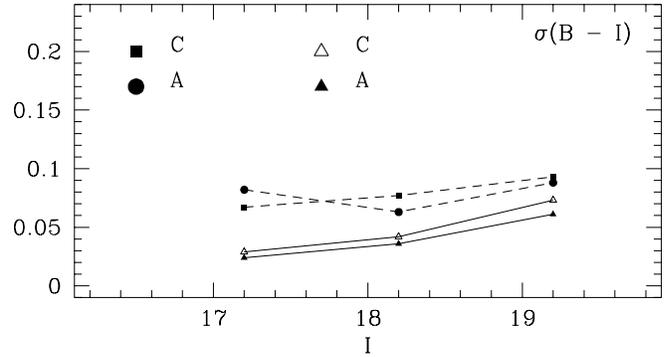,width=8.8cm,bbllx=20pt,bblly=150pt,bburx=565pt,bbury=450pt}
 \vspace{0cm}
 \caption[]{
 Plot of the measured color scatter $\sigma_{(B-I)}$\ of the {\it
 principal} RGB component in our fields A and C (circles), compared
 with instrumental dispersions (triangles).
 \label{f_cfsigmas} }
 \end{figure}

 \tablecaption{Observed and instrumental color dispersions $\sigma_{(B-I)}$ for 
 the main component of the RGB color distributions in fields A and C. 
  \label{t_sigmas} }
 \begin{planotable}{ccccccc}
 \tablehead{
 \colhead{$I$} &
 \colhead{$\sigma_{\rm A}$} &
 \colhead{$\sigma_{\rm C}$} &
 \colhead{$\sigma_{\rm A}{\rm(instr)}$} &
 \colhead{$\sigma_{\rm C}{\rm(instr)}$} &
 \colhead{$\sigma_{0,{\rm A}}$} &
 \colhead{$\sigma_{0,{\rm C}}$} 
 }
 \startdata
 %
 %
  17.20 & 0.082 & 0.067 & 0.024 & 0.029  & 0.078 & 0.060  \\  
  18.20 & 0.063 & 0.077 & 0.036 & 0.042  & 0.052 & 0.065  \\  
  19.20 & 0.088 & 0.093 & 0.061 & 0.073  & 0.063 & 0.058  \\ 
 \end{planotable}

One of the main results of this paper, made possible by the size of
our stellar sample and photometric baseline, is a detailed analysis of
the color distribution function (\abbrev{\cdf}) of the red giant stars
in Fornax. Fig.~\ref{twoGauss} shows the distribution of the ($B-I$)
color residuals about a preliminary fiducial sequence, in the
magnitude range $17.7 \leq I \leq 18.7$\ ($-3< M_I <-2$), for the
inner and outer field.
%
%
While these histograms confirm the well-known wide color range of the RGB stars 
in 
Fornax (\eg Buonanno \etal \cite{b85}; Sagar et al. \cite{sagar}; Grebel \etal 
\cite{greb+94}; Beauchamp \etal \cite{beauchamp}), they also show that the color 
distributions cannot merely be described using a single ``color dispersion''. 
Rather, the \cdf\ is more appropriately described as roughly bimodal, showing a 
principal peak and a bluer component extending to 
$\Delta (B-I)\simeq -0.4$. 
This color distribution function is quite well modeled by the sum of
two Gaussians. The main component of the distribution represents the bulk of
the red giant population, a mix of old and (mostly) intermediate-age
stars.  The secondary peak is centered at about $\Delta_{B-I}=-0.20$.

%
On the other hand, we notice a relatively well-defined cutoff on the red side of 
the 
RGB, indicating the lack of any significant metal-rich component similar to the 
stellar population of 47~Tuc, or even less metal-rich if we assume a mean age 
younger than that of Milky Way globular clusters. This absence sets an important 
constraint for modeling the chemical enrichment of the Fornax dwarf. 
Both components are wider than accounted for by instrumental errors.
The dispersions of the two components, in the luminosity range
$-3 < M_I < -2$, are 
$\sigma_{B-I}^a = 0.063$\ (central peak) and $\sigma_{B-I}^b = 0.088$\ 
(blue component) in the inner region, and 
$\sigma_{B-I}^a = 0.077$, $\sigma_{B-I}^b = 0.099$\ in the outer field. 
Table~\ref{t_sigmas} gives the observed dispersions for the main component
of the Fornax field population for the inner and outer region in 3 magnitude 
intervals. 
Also given in Table~\ref{t_sigmas} are the instrumental errors $\sigma(B-I)$\ 
obtained by fitting a Gaussian to the color residuals of artificial stars, 
exactly in the 
same way as for the real data. 
%
The observed and instrumental dispersions are also compared in 
Fig.~\ref{f_cfsigmas}. 
%
%
The intrinsic ($B-I$) color dispersions, calculated as the quadratic difference 
between the observed and the instrumental scatter, are given in the last two 
columns 
of Table~\ref{t_sigmas}. In the luminosity interval 
$ -3 < M_I < -2$\  the intrinsic color spread of the main RGB population is 
$\sigma_0 (B-I) = 0.06 \pm 0.01$\ mag. 
%
Using again the color-color relations for the Fornax red giants to convert ($B-
I$) 
color spreads into equivalent dispersions in ($V-I$), and the calibration of RGB 
color 
shifts as a function of metal abundance variations, we obtained a metallicity 
spread 
$\sigma_{\rm [Fe/H]} = 0.12 \pm 0.02$\ dex for the dominant field population. 
%
%

Then the ($2\sigma$) metallicity range for the bulk population of Fornax would 
be 
approximately 
$-1.65 < {\rm [Fe/H]} < -1.15$, or 
$-1.25 < {\rm [Fe/H]} < -0.75$\ if a correction for the mean age is applied.
This intrinsic metallicity range is significantly lower than the abundance 
spread 
quoted by most previous studies for the red giant branch as a whole.  
According to Beauchamp et al. (\cite{beauchamp}), the total range in [Fe/H]
is 0.8 dex, comparable to that found by Sagar et al. (\cite{sagar}) and Grebel 
et al. 
(\cite{greb+94}). A smaller spread ($\lesssim 0.1$~dex) was found by Geisler 
(\cite{geisler}).  This discrepancy probably results from the coarser 
metallicity resolution of the colors employed in past studies, with the notable exception of the Washington colors of Geisler (\cite{geisler}).
We conclude that a small abundance spread seems in fact more
appropriate to describe the intermediate-age field population in
Fornax.  While the metallicity dispersion given above is comparable to
that of Leo~I (\eg Gallart et al. \cite{gall+99}), it appears to be
smaller than the abundance spread found in the majority of dwarf
spheroidal galaxies (Da Costa
\cite{dacosta_winter}; 
Mateo \cite{mate98}). In a few cases, wide range in metallicity has been 
confirmed by low- and high-dispersion spectroscopy (\eg Cot\'e et al. \cite{cote+99}; Shetrone 
et al. \cite{shet+98}). The mean value of the metallicity spread for Galactic 
dSph and satellites of M\,31 is $0.37\pm0.03$\ dex (Cot\'e et al. 
\cite{cote+99}).
Had we considered our Fornax RGB color distribution as a whole, we would have obtained a metallicity spread of the same order ($-2.0 < {\rm [Fe/H]} < -0.7$, $\pm2\sigma$\ range), in good agreement with previous studies.

Since age is known to affect the RGB color, [Fe/H] dispersions derived
by the width of the giant branch should be taken with caution in view
of a possible contribution of an age spread. As argued above, an age
range of the order 5 Gyr (which is that of stars making up the main
RGB) is sufficient to mimic a metallicity range of $\pm 0.1$\
dex. Thus we might assume that the width of the RGB main component is
entirely due to the age spread of its populations. The situation is
more complex, though, and the effects of a metallicity and age range
on the color distribution depend on the details of the star formation
and chemical enrichment history. Successive stellar generations are
expected to be progressively more metal-enriched, so that younger
stars (implying a bluer RGB) will generally have higher metal
abundance (leading to a redder RGB). The two effects -- of a younger
age and higher metallicity -- will act in opposite directions, and may
even compensate each other as it appears to be the case for Carina
(Smecker-Hane et al. \cite{smec+94}). Similarly to Carina, the
abundance spread we have found for the dominant population of Fornax
may represent a {\it lower limit} (see also Paper~II; Gallart et
al. \cite{gall+99}; for similar considerations for other dSph's). This
issue shall be more quantitatively investigated in a following paper.

We return now to discuss the nature of the population making up the blue tail of 
the 
\cdf, which is until now far from established. 
Qualitative examination of the \cmd's is not sufficient to establish whether the 
blue 
tail of the \cdf\ represents an old, \hbox{metal-poor} population, or is made up 
of 
young red AGB stars. However, we will show in Sect.~\ref{sec_popratios} that 
there 
is definite evidence that the bluer RGB stars are old and metal-poor, which 
implies 
that the extended color distribution shown in Fig.~\ref{twoGauss} can be 
interpreted 
as a metallicity distribution. In conclusion, a model involving two populations 
seems 
to provide a good description of the star content of the Fornax dSph, with the 
older 
population having [Fe/H]$= -1.82$\ with a dispersion of 0.20 dex, and the 
dominant, 
intermediate-age population with [Fe/H]$\approx \corfeh \pm \errfeh$. 
Our large-field data confidently rule out the presence of a distinct metal-rich 
population with abundance comparable to that of 47~Tuc, even accounting for a 
mean age of 5 Gyr for the Fornax bulk population.
%

This complexity is common to most of the other dwarf spheroidals.
For example, two distinct star formation epochs have been recently 
revealed in Sculptor by Majewski et al. (\cite{maje+99}). In this galaxy, a detailed analysis of the RGB morphology showed the presence of two distinct RGB bumps consistent with the presence of a metal-poor population of [Fe/H]$\sim -2.3$, and a population of [Fe/H]$\sim -1.5$.
Also the recent study of the star formation history of Leo~I by
Gallart et al. (\cite{gall+99b}) indicate that most of the star
formation activity (80\%) occurred between 7 and 1 Gyr (mean 4 Gyr)
while the contribution of the older phase was small.
A wide metallicity range and a composite population, although with a
higher mean abundance, has also been inferred in the Sagittarius dSph,
a galaxy similar in many respects to Fornax (Bellazzini et
al. \cite{bell+99}).  Also, the metallicity distribution of stars in
the small elliptical M\,32 shows a metal-rich peak ([Fe/H]$\simeq
-0.2$) with a low-metallicity tail extending to about [Fe/H]$\sim-1.5$
(Grillmair et al. \cite{gril+96}).
It is also interesting to note the analogy with the extremely broad
metallicity range found in the halo of the nearby elliptical NGC 5128
(Harris et al. \cite{ghar+98}), where the shape of the metallicity distribution suggested a two-phase {\em in situ} model.


\subsection{Population gradients} 
\label{sec_popratios} 



\begin{figure}
\vspace{0cm}     
\hspace{0cm}\psfig{figure=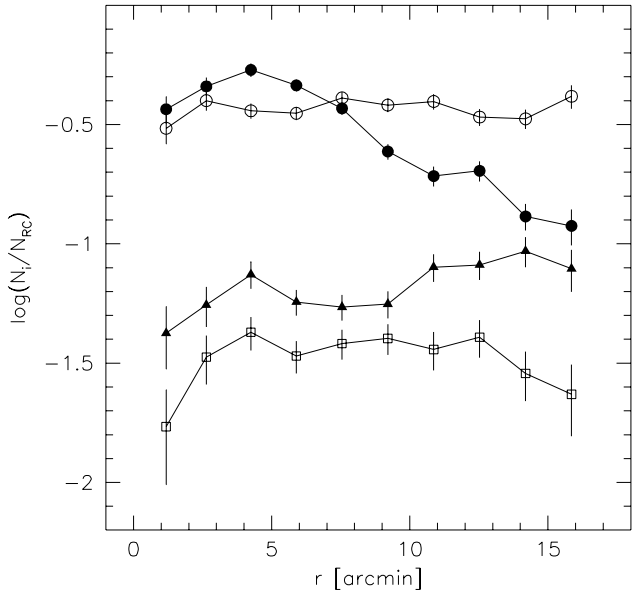,width=8.8cm}
\vspace{0cm}
\caption[]{Radial trends in the fraction of young main sequence stars 
(filled circles), HB stars (triangles), stars on the blue-RGB
(squares) and RGB (open circles) relative to the number of red clump
stars. The logarithm of the ratios is plotted against the effective
distance from the Fornax center.
 }
\label{f_popratios}
\end{figure}




\tablecaption{
Star counts normalized to 100 stars in the red clump
\label{t_popratios} }
\begin{planotable}{rrrrrr}
\tablehead{
\colhead{$r\arcmin$} & 
\colhead{HB} &
\colhead{BL} &
\colhead{yMS} &
\colhead{$b$RGB} &
\colhead{RGB} 
}
\startdata 
      1.2 &      4.2 &      7.0 &     36.6 &      1.7 &     30.5\nl
      2.6 &      5.5 &      6.3 &     45.7 &      3.3 &     39.8\nl
      4.2 &      7.4 &      9.3 &     53.6 &      4.3 &     36.1\nl
      5.9 &      5.7 &      7.4 &     46.2 &      3.4 &     35.2\nl
      7.5 &      5.4 &      5.9 &     37.0 &      3.8 &     40.9\nl
      9.2 &      5.6 &      5.2 &     24.4 &      4.0 &     38.1\nl
     10.9 &      8.0 &      6.0 &     19.2 &      3.6 &     39.5\nl
     12.5 &      8.2 &      6.1 &     20.2 &      4.1 &     34.0\nl
     14.2 &      9.3 &      4.4 &     13.0 &      2.9 &     33.4\nl
     15.9 &      7.9 &      5.7 &     11.9 &      2.3 &     41.5\nl
\end{planotable}


A comparison of the color-magnitude diagram in the different regions
in this study provided important clues regarding the origin of the
stellar populations in Fornax, and in particular on the nature of its
complex red giant branch.
Were the bluer RGB stars old and metal-poor, one would expect a higher
fraction of them in the outer fields, on the basis of the population
gradient detected by SHS98. 
Clearly the opposite finding, i.e. a larger RGB blue tail in the inner
regions, would indicate a connection to the more recent bursts of star
formation.

In order to measure the radial gradient in the stellar populations in
Fornax, stars in different evolutionary phases were counted separately
in different radial bins. The \cmd\ regions chosen for counts
include the red clump, the red part of the \oldhb, the blue-loop
helium-burning stars (BL), the young main sequence (yMS), and the red
giants (those in the mainstream giant branch, RGB, and in the bluer
component, $b$RGB).

The reader is referred to the boxes outlined in Fig.~\ref{1cmdBVI}.
%
The results of star counts are presented in Table~\ref{t_popratios},
where we list the effective galactocentric distance and the
{percentage} of stars in all the \cmd\ regions relative to the number
of RC stars.

 The fraction of young main sequence, old HB, blue-RGB and mainstream
 RGB stars are also plotted on a logarithmic scale in
 Fig.~\ref{f_popratios}. As previously noticed by SHS98, the young
 stars are more centrally concentrated than the dominant
 intermediate age component, indicating that recent star formation
 took place preferentially in the central regions.  The counts on the
 RGB as expected follow those of RC stars. 
%
%
Conversely, the HB stars are
 preferentially found in the outer regions.

Most importantly, the bluer RGB stars {\it closely follow the radial
trend of the horizontal-branch stars} (Fig.~\ref{f_popratios}). This
result unambiguously demonstrates that the sparse sequence on the blue
side of the Fornax RGB belongs to the {\it old and metal-poor}
population ($\gtrsim 10$\ Gyr) along with the old-HB stars and RR
Lyrae variables.

\section{Summary and conclusions}  \label{sec_conclu}

We have presented a large area study of the field population in the
Fornax dwarf spheroidal galaxy, based on $BVI$\ data for about 40000
stars. The size of our sample, together with the wide photometric
baseline employed in this work, provide new information on the stellar
content of Fornax.

One of the most distinctive features in our diagrams is the
conspicuous young main sequence. In this paper we have shown that the
plume of stars just above the red clump is made up of intermediate
mass stars (2.4--2.9 \msol) burning helium in the core, counterparts
of the young main sequence stars in the age range 0.3--0.4 Gyr.  The
comparison with isochrones suggests us that these blue-loop stars must
be as metal-rich as [Fe/H]$\sim-0.7$, which represents an important
constraint for the metal enrichment history in Fornax.

An extended upper AGB tail and a prominent red HB clump testify the
presence of a dominant intermediate-age population in the age range
2-10 Gyr, corresponding to 0.9--1.4 \msol stars. From the difference
in the mean ($V-I$) colors of the red clump and the RGB at the same
luminosity, we have estimated a mean age $5.4\pm1.7$\ for the bulk of
the intermediate-age population in Fornax, in agreement with the
morphology of the MS turnoffs in WFPC2 color-magnitude diagrams
(Buonanno et al. \cite{buon+99}). This suggests that the location of
the red HB clump may indeed prove to be a useful age indicator for
distant LG galaxies.

About 0.2 mag below the red clump, an extended HB is indicative of an
old population. In particular, our data point to the presence of blue
HB stars in the outer regions. Together with previous detection of RR
Lyrae, this provides evidence for a minority field population that is
as old and metal-poor as that in the Fornax globular clusters. The
Fornax dSph clearly started forming stars in a halo nearly at the same
epoch when most of its surrounding clusters were formed.

Evolutionary phases that gave barely discernible features in small
field observations are easily measurable in our color-magnitude
diagrams.  We could reliably measure the AGB bump, a small clump
produced by a clustering of stars at the base of the AGB, at
$M_V\simeq-0.4$. Measurements of such minor evolutionary features may
provide useful tests of stellar models for stars of different masses
and metallicities.

The sharp cutoff in the luminosity function of Fornax has been used to
estimate its distance using the RGB tip method. The corrected distance
modulus of Fornax, $(m-M)_0 = 20.70 \pm 0.12$, agrees with previous
determinations.
This estimate is confirmed by the mean level of old horizontal-branch
stars. By measuring the average magnitude of the red HB (distinct from
the red clump) we estimated a distance modulus $(m-M)_0 = 20.76 \pm
0.12$\ on the distance scale of Lee et al. (\cite{ywlee+90}).

Fornax, as many other dSph, has been known for a long time to have a
wide RGB color distribution. The ``color scatter'' has been usually
taken to represent an abundance spread. We have analyzed in detail the
color distribution of the red giant stars across the fiducial line,
and found that it is reasonably well fitted by a two-component
model. This approximately bimodal distribution is remarkably similar
in all fields. About 70\% of the red giants belong to an
intermediate-age RGB component which
is itself wider than expected from instrumental errors. By comparing
the bulk of the Fornax RGB with the ridge lines of standard globular
clusters, we have estimated a mean metallicity [Fe/H]$=\feh \pm
\errfeh$. This nominal value should be corrected for the age
difference between the Fornax population and the Milky Way globulars.
Accounting for an age difference of 10 Gyr, we find an {\it
age-corrected} mean metallicity [Fe/H]$ = \corfeh \pm \errfeh$ for the
dominant intermediate-age population of Fornax. Interestingly, this is
also the metallicity found for Sagittarius, the nearest Milky Way dSph
satellite that has luminosity comparable to that of Fornax.
The {\it intrinsic} color scatter of stars in the RGB main component is 
$\sigma_{(B-I)}=0.06\pm0.01$\ mag implying a relatively modest metallicity 
spread 
$\sigma_{\rm [Fe/H]} = 0.12 \pm 0.02$\ dex. Then the ($2\sigma$) metallicity 
range 
for the bulk population of Fornax is 
$-1.25 < {\rm [Fe/H]} < -0.75$\ if a correction for the mean age is applied.
The secondary component or ``bluer tail'' is also quite broad. 
In principle, these bluer stars could be either young or old and metal-poor.

Star counts of different subpopulations at various locations confirm
and extend the evidence for radial population gradients emerged in
previous studies. Recent star formation is clearly concentrated in the
central regions, though with some degree of asymmetry (\eg SHS98).
Old stars are more easily seen in the outer fields.
A blue HB population can be noticed in our outermost field, coming from the minority old, metal-poor field component.
The stars populating the blue side of the wide RGB closely follow the 
spatial distribution of the old-HB stars. This is perhaps our most important 
finding, 
since it demonstrates that the bluer RGB stars are themselves old and metal-
poor, and 
clearly establishes the nature of the wide RGB of Fornax. 
Thus the roughly bimodal color distribution can be interpreted as a metallicity 
distribution, implying that the bulk of the Fornax galaxy was built during two 
rather 
distinct star-forming epochs. The older population has 
[Fe/H]$\approx -1.8$\ dex with ($\pm 2\sigma$) and an wide abundance range 
$-2.2 < {\rm [Fe/H]}< -1.4$. 

The emerging picture is one in which the evolution of Fornax is characterized by 
two 
major star formation epochs, each consisting of many episodes. The first episode 
took place at an early epoch, being presumably coeval to the birth of the old 
galactic 
globular clusters, from metal-poor gas. After a relatively quiescent period, 
Fornax 
formed the bulk of stellar populations between 7 and 2.5 Gyr ago from the 
pre-enriched gas. Star formation continued at a lower rate in the central 
regions until 
as recently as 10$^8$\ yr ago. The modest internal abundance spread found in 
each 
main population seen in the metallicity distribution, and the different 
metallicities of 
populations of different age, trace the progressive metal enrichment and 
represent the 
basis for an age-metallicity relation in Fornax. 
The constraints found in this paper provide the physical input for a 
quantitative 
analysis of the star formation and chemical enrichment history of Fornax, which 
will 
be done in a forthcoming study using the methods of stellar population 
synthesis. 

\begin{acknowledgements}
We thank L. Girardi for useful discussions and for kindly providing us
with unpublished theoretical red clump colors.
Dr. P.B. Stetson is thanked for helpful comments on the manuscript.
I.~S.  acknowledges support from ANTARES, an astrophysics network
funded by the HCM program of the European Community.
\end{acknowledgements}



\end{document}